\documentclass[aps,prd,twocolumn,groupedaddress,nofootinbib,showpacs]{revtex4}%

\usepackage{graphicx}

\usepackage{amsmath}
\usepackage{amssymb}
\usepackage{multirow}
\usepackage{bm}
\usepackage{color}

\usepackage[hypertex]{hyperref}

\def\be{\begin{equation}}
\def\ee{\end{equation}}
\def\ba{\begin{eqnarray}}
\def\ea{\end{eqnarray}}

\def\chic{\chi_{c1}'}
\def\ochic{{\langle\mathcal{O}^{\chic}_n\rangle}}

\def\NO{\nonumber}

\begin{document}



\title{\Large $X(3872)$ and its production at hadron colliders}
\author{Ce Meng$^a$}
\email{mengce75@pku.edu.cn}
\author{Hao Han$^a$}
\email{unknowndear@163.com}
\author{Kuang-Ta Chao$^{a,b,c}$}
\email{ktchao@pku.edu.cn}

\affiliation{ {\footnotesize (a)~School of Physics and State Key
Laboratory of Nuclear Physics and Technology, Peking University,
Beijing 100871, China}\\
{\footnotesize (b)~Collaborative Innovation Center of Quantum Matter, Beijing 100871, China}\\
{\footnotesize (c)~Center for High Energy Physics, Peking
University, Beijing 100871, China}
}
\date{\today}

\begin{abstract}
We evaluate the production cross sections of $X(3872)$ at the LHC and Tevatron at NLO in $\alpha_s$ in NRQCD by assuming that the short-distance production proceeds dominantly through its $\chi_{c1}'$ component in our $\chi_{c1}'\mbox{-}D^0\bar{D}^{*0}$ mixing model for $X(3872)$. The outcomes of the fits to the CMS $p_T$ distribution can well account for the recent ATLAS data in a much larger range of transverse momenta ($10~\mbox{GeV}<p_T<70~\mbox{GeV}$), and the CDF total cross section data, and are also consistent with the value of $k=Z_{c\bar c}\cdot Br(X\to J/\psi\pi^+\pi^-)$ constrained by the $B$-meson decay data. 
For LHCb the predicted X(3872) total cross section is larger than the data by a factor of 2, which is due to the problem of the fixed-order NRQCD calculation that may not be applicable for the region with small $p_T$ ($p_T\sim 5 ~\mbox{GeV}$) and large forward rapidity $(2.5<y<4.5)$.
In comparison, the prediction of molecule production mechanism for $X(3872)$ is inconsistent with both $p_T$ distributions and total cross sections of CMS and ATLAS, and the total cross section of CDF.
\end{abstract}


\maketitle


\section{Introduction}
The hidden-charm state $X(3872)$ was first discovered by the Belle Collaboration in the $J/\psi\pi^+\pi^-$ invariant mass spectrum of $B^+\to J/\psi\pi^+\pi^-K^+$ decay\cite{Belle03}, and confirmed by the CDF\cite{CDF04}, D0\cite{D004} and BaBar\cite{BaBar05} collaborations soon after. The closeness of the mass $m_X=3871.68\pm0.17$~MeV~\cite{PDG2012} to the $D^0\bar{D}^{*0}$ threshold led many authors to speculate that the $X$ is a $D^0\bar{D}^{*0}$ molecule~\cite{Melecule04}. (For recent reviews of $X(3872)$, see Ref.~\cite{Brambilla})

However, with the tiny binding energy $E_b=(m_{D^0}+m_{\bar{D}^{*0}})-m_X=0.142\pm0.220$~MeV\cite{Tomaradze2012-X3872-Eb}, it is difficult to imagine that such a loosely bound state can have a large prompt production rate in $p\bar{p}$ collisions. In addition, D0 found\cite{D004} that the behavior of $X(3872)$ production in $p\bar{p}$ collision is very similar to that of $\psi'$, such as the cross-section $p_T$ distribution. In view of these, in 2005 we proposed that the $X(3872)$ is a mixed state of $\chi_{c1}(2P)$ ($\chi_{c1}'$) and $D^0\bar{D}^{*0}$ due to coupled channel effects\cite{Meng-BtoX3872}. The two components are both substantial and may play different roles in the dynamics of $X(3872)$: the short-distance (the $B$- and hadro-) production and the quark annihilation decays of $X$ proceed dominantly through the $\chi_{c1}'$ component; while the $D^0\bar{D}^{*0}$ component is mainly in charge of the hadronic decays of $X(3872)$ into $DD\pi,DD\gamma$ as well as $J/\psi\rho$ and $J/\psi\omega$.

Based on the calculation for the ratio $\frac{Br(B\to \chi_{c1}'K)}{Br(B\to \chi_{c1}K)}$, we estimate that $Br(B\to \chi_{c1}'K)=(2\mbox{-}4)\times10^{-4}$\cite{Meng-BtoX3872}, which is consistent with the experimental constraints on $Br(B\to X(3872)K)$  renormalized by $Z_{c\bar{c}}$, where $Z_{c\bar{c}}$ is the probability of the $\chi_{c1}'$ component in the $X(3872)$.

We also calculate the $X$ decays into $J/\psi\rho$ and $J/\psi\omega$ through rescattering of the intermediate states, i.e. the $D^0\bar{D}^{*0}$ component, and this may account for the observed large isospin violation\cite{Meng-X3872-rescattering}.

The closeness of the mass of $X(3872)$ to the $D^0\bar{D}^{*0}$ threshold may be explained by the S-wave coupling between the $J^{PC}=1^{++}$ $c\bar c$ component ($\chi_{c1}'$) and the $D^0\bar{D}^{*0}$ component, which induces a sharp spectral density and lowers the "bare" mass of $\chi_{c1}'$ towards the $D^0\bar{D}^{*0}$ threshold\cite{X3872-Mass-CoupledChannel}.

A crucial test for the need of the $\chi_{c1}'$ component in $X(3872)$ was suggested in our 2005 paper~\cite{Meng-BtoX3872}, that the E1 transition rate of $\chi_{c1}'\to \psi'\gamma$ is evidently larger than that of $\chi_{c1}'\to J/\psi\gamma$. This was supported later by the BaBar measurement that the observed ratio $R_{\gamma}=\frac{Br(X(3872)\to \psi'\gamma)}{Br(X(3872)\to J/\psi\gamma)}=(3.4\pm 1.4)$~\cite{Aubert:2008ae}.
In contrast, the molecule model predicted the ratio $R_{\gamma}$ to be much less than 1. Moreover, the observed large ratio of $\frac{Br(X(3872)\to\psi'\gamma)}{Br(X(3872)\to J/\psi\rho(\omega))}\sim 1$, is hard to explain in the molecule model.

As for the prompt production of $X$ in $p\bar{p}$ collision, Suzuki\cite{Suzuki2005} also pointed out that the molecule assumption can hardly be consistent with the CDF measurement, based on an estimation of the wave function at the origin for
the molecule. An explicit calculation given by Bignamini {\it et al.}\cite{Bignamini2009} set the upper bound of the cross section of $X$ as a $D^0\bar{D}^{*0}$ molecule to be $0.085$ nb, which is smaller than the experimental value given by
\be
\sigma(p\bar{p}\to X)\times Br(X\to J/\psi\pi^+\pi^-)=3.1\pm0.7~\mbox{nb}.
\label{CDF-X3872}\ee
by about 2 orders of magnitude.

Artoisenet and Braaten~\cite{Braaten2010-X3872HadroPro} suggested that the $D\bar{D}^*$ rescattering effects could enhance the molecule production cross section to values consistent with the CDF data if one chooses the upper bound of the relative momentum of $D\bar{D}^*$ rescattering to be $3m_{\pi}$ or more. With NRQCD factorization ~\cite{BBL05-NRQCD} at leading order (LO) and with the matrix element extracted from the CDF data of $X(3872)$, they also made predictions for the molecule production cross section at the LHC, which is about 3 to 4 times larger than the recent LHCb~\cite{LHCb12-CroSec} and CMS~\cite{Chatrchyan:2013cld} data.

In our view, with the $\chi_{c1}'\mbox{-}D^0\bar{D}^{*0}$ mixing model, one can naturally understand the prompt production of $X(3872)$ at the Tevatron and LHC: the $X(3872)$ production mainly proceeds via the $\chi_{c1}'$ component and shares similar behavior to that of $\psi'$. In this paper, we will study the prompt production of $X(3872)$  at next-to-leading order (NLO) in NRQCD, and compare our results with the molecule model on both the total cross sections and $p_T$ distributions. We will also discuss the constraints on the value of $Z_{c\bar{c}}$ indicated by our calculation.

\section{Prompt~X(3872)~production~in~NRQCD}

In the $\chic$-dominant production mechanism, the inclusive cross section of $X$ in the $J/\psi\pi^+\pi^-$ mode can be expressed as
\be
d\sigma(pp\to X(J/\psi\pi^+\pi^-))
=d\sigma(pp\to \chi_{c1}')\cdot k,
\label{Factorization-X3872}\ee
where $p$ is either a proton or an antiproton, and $k=Z_{c\bar{c}}\cdot Br_0$ with $Br_0=Br(X\to J/\psi\pi^+\pi^-)$. The feed-down contributions from higher charmonia (e.g. $\psi(3S)$ are negligible for the prompt production of $X(3872)/\chi_{c1}'$, so here "prompt" is almost equal to "direct", and the cross section of $\chi_{c1}'$ in Eq.~(\ref{Factorization-X3872}) can be evaluated in NRQCD factorization, which is given by
\begin{eqnarray}
\label{Factorization-chic2P}
&d\sigma(pp\rightarrow\chic)=\sum_{n}
d\hat{\sigma}((c\bar{c})_n)\frac{\ochic}{m_c^{2L_n}}\\
&=\sum_{i,j,n} \int dx_1dx_2\, G_{i/p}G_{j/p}
d\hat{\sigma}(ij\rightarrow (c\bar{c})_n)\ochic,
\NO
\end{eqnarray}
where $G_{i,j/p}$ are the parton distribution functions (PDFs) of $p$, and the indices $i, j$ run over all
the partonic species. The matrix element $\ochic$ is marked by "$n$", which denotes the color, spin and angular momentum of the intermediate $c\bar{c}$ pair. Here we will evaluate the cross section at NLO in $\alpha_s$ and at LO in $v$ (the relative velocity of $c\bar{c}$ in the rest frame of $\chic$); therefore, only $n = ^3\!P_1^{[1]}$ and $^3\!S_1^{[8]} $ are present here.

Since $\chi_{c1}(1P)$ and its radial excitation $\chic$ share the same quantum number, they are physically alike. For the purpose of comparison, we follow the conventions of Ref.~\cite{Chao:chic}, where similar calculations are done for $\chi_{c1}$, to define the color-singlet matrix element as
\be
\langle\mathcal{O}^{\chic}(^3\!P_1^{[1]})\rangle=\frac{9}{4\pi}|R^{\prime}_{2P}(0)|^2,
\label{3P11}\ee
where $R^{\prime}_{2P}(0)$ is the derivative of the $\chic$ radial wave function at the origin, and the definition in Eq.~(\ref{3P11}) is different from that in Ref.~\cite{BBL05-NRQCD} by a factor of $1/(2N_c)$. Similarly, we parameterize the color-octet matrix element by the ratio
\be
r=m_c^2\langle\mathcal{O}^{\chic}(^3\!S_1^{[8]})\rangle/\langle\mathcal{O}^{\chic}(^3\!P_1^{[1]})\rangle
\label{r}\ee
as in Ref.~\cite{Chao:chic}. Thus, the cross section in Eq.~(\ref{Factorization-X3872}) will be a function of the parameters $r$, $k$, and $|R^{\prime}_{2P}(0)|^2$. For  simplification, we fix the value of the wave functions
\be
|R^{\prime}_{2P}(0)|^2=|R^{\prime}_{1P}(0)|^2=0.075~\mbox{GeV}^5,
\label{R0}\ee
where in the second equality, the value for $\chi_c(1P)$ is chosen from the B-T-type potential model calculation~\cite{eichten} and was successfully used to evaluate the cross sections of $\chi_{c1,2}$ to account for the LHCb measurements~\cite{LHCb12-chic1P}, and the first equality $|R^{\prime}_{2P}(0)|^2=|R^{\prime}_{1P}(0)|^2$ is assumed based on various potential model calculations~\cite{eichten}, and this assumption has been adopted in our previous evaluation of the ratio $\frac{Br(B\to \chi_{c1}'K)}{Br(B\to \chi_{c1}K)}$~\cite{Meng-BtoX3872}. So it will be convenient to compare the results of $X$ production in $pp$ collision with that in B decay. Actually, any change of $|R^{\prime}_{2P}(0)|^2$ can be simply compensated by a corresponding change of $k$, and we will return to this point later.

As for the numerical calculation, we choose the same input parameters as Ref.\cite{Chao:chic}.
We use the CTEQ6L1 and CTEQ6M PDFs~\cite{Whalley05-CTEQ6M} for LO and NLO calculations respectively. The charm quark mass
is set to be $m_c = 1.5$~GeV; meanwhile, the renormalization, factorization and NRQCD scales are
$\mu_r = \mu_f =m_T\equiv \sqrt{p_T^2+4m_c^2}$ and $\mu_{\Lambda} = m_c$. To estimate theoretical uncertainties, we vary $\mu_r$ and $\mu_f$ from $m_T/2$ to
$2m_T$ and choose $m_c = 1.5\pm0.1$ GeV. We refer other details of the calculations to Ref.\cite{Chao:chic}.

One should note that the $D^0\bar{D}^{*0}$ molecule can also be produced in $pp$ collision through the short-distance $c\bar{c}$ pair production with small relative velocity $v$. Therefore, the cross section can be factorized similarly to that in Eq.~(\ref{Factorization-chic2P}) and given by~\cite{Braaten2010-X3872HadroPro}
\be
\sigma(D^0\bar{D}^{*0})=\hat{\sigma}(^3\!S_1^{[1]})\langle {\cal O}^{D\bar{D}^*}(^3\!S_1^{[1]}) \rangle+\hat{\sigma}(^3\!S_1^{[8]})\langle {\cal O}^{D\bar{D}^*}(^3\!S_1^{[8]}) \rangle,
\label{Factorization-DD*}\ee
where the PDFs have been integrated out in the coefficients $\hat{\sigma}(^3\!S_1^{[1,8]})$, and the NRQCD velocity scaling rule~\cite{BBL05-NRQCD} and heavy quark spin symmetry~\cite{Voloshin04-HQSS} have been adopted to truncate the formula to $^3\!S_1^{[1,8]}$ terms. The coefficients $\hat{\sigma}(^3\!S_1^{[1,8]})$ are the same as those in the charmonium cross section. We evaluate them at NLO in the same scheme mentioned above, and we find that the ratio $\hat{\sigma}(^3\!S_1^{[1]})/\hat{\sigma}(^3\!S_1^{[8]})$ is about $5.3\times10^{-4}$ for the CDF with $p_T>5$ GeV, and $1.5\times10^{-4}$ for the CMS with $p_T>10$ GeV (the coefficient $\hat{\sigma}(^3\!S_1^{[1]})$ has been divided by $2N_c$ to match the convention in Ref.~\cite{Braaten2010-X3872HadroPro}). Note that the matrix elements in Eq.~(\ref{Factorization-DD*}) are of the same order, and thus the color-singlet contributions can be neglected.

Furthermore, the matrix elements $\langle {\cal O}^{D\bar{D}^*}(^3\!S_1^{[1,8]}) \rangle$ for a loosely bound state should be much smaller than those of charmonia, which can be justified by the calculations of Refs.~\cite{Suzuki2005,Bignamini2009}; thus, we neglect the contributions from Eq.~(\ref{Factorization-DD*}) in our mixing model. However,  Artoisenet and Braaten~\cite{Braaten2010-X3872HadroPro} argued that the $D\bar{D}^*/\bar{D}D^*$ rescattering effects could enhance the molecule matrix elements to be consistent with the CDF data of total cross section for $X(3872)$. Anyway, the two models are different in different combinations of channels in Eqs.~(\ref{Factorization-chic2P}) and (\ref{Factorization-DD*}). This result is thanks to the cross section $p_T$ distribution of $X(3872)$ measured by the CMS Collaboration~\cite{Chatrchyan:2013cld}, which allows us to compare the two different combinations in the two models.

\begin{figure}
\includegraphics[width=7cm]{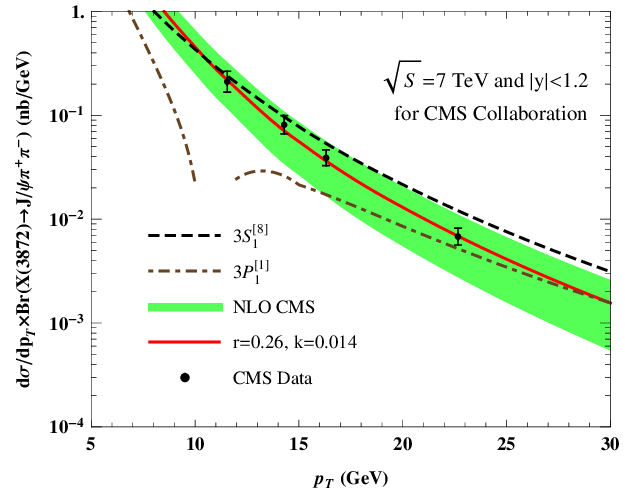}
\caption{\label{fig:diagrams} The $p_T$ distribution of the prompt production cross section of $X(3872)$. The CMS data are taken from Ref.~\cite{Chatrchyan:2013cld}. The theoretical curves are obtained from the two-parameter fits. The dashed, dot-dashed and solid lines represent the $^3\!S_1^{[8]}$, $^3\!P_1^{[1]}$, and total contributions, respectively. The green bands denote the uncertainties of the total results.}
\end{figure}

Using formulas (\ref{Factorization-X3872}) and (\ref{Factorization-chic2P}), we fit the CMS $p_T$ data ($\sqrt{S}=7~\mbox{TeV},~|y|<1.2$)~\cite{Chatrchyan:2013cld} by minimizing the $\chi^2$, and the results are shown in Fig.~1 with the outcomes
\be
r=0.26\pm0.07,~~~~k=0.014\pm0.006,
\label{r&k}\ee
where the central values correspond to $\chi^2/2=0.26$, and the large error-bars, indicated by the broad band in Fig.~1, are due to the insensitivity of the $p_T$ distribution to the parameter $r$ in the range 0.20-0.40. Nevertheless, one can see that the central value of $r$ in Eq.~(\ref{r&k}) is almost the same as that in Ref.~\cite{Chao:chic}; i.e., $r=0.27$ for $\chi_{c1}(1P)$. This may imply that $X(3872)$ can be produced through its $\chic$ component at short distances. In comparison, the $p_T$ behavior of the $^3\!S_1^{[8]}$ channel is also shown solely in Fig.~1, which can hardly explain the data. Thus, the molecule production mechanism in Eq.~(\ref{Factorization-DD*}), where the contribution from the $^3\!S_1^{[1]}$ can be neglected as mentioned above, is disfavored by the CMS data. More explicitly, we use (\ref{Factorization-DD*}) to fit the $p_T$ distribution and get
\be
\langle {\cal O}^{D\bar{D}^*}(^3\!S_1^{[8]}) \rangle Br_0=6.0\times 10^{-5}~\mbox{GeV}^3,~~\chi^2/3=1.03.
\label{fit-molecule}\ee

For the CDF window ($\sqrt{S}=1.96~\mbox{TeV},~|y|<0.6,~p_T>5~\mbox{GeV}$), using the central values in Eq.~(\ref{r&k}), we predict the total cross section to be
\be
\sigma^{th}_{CDF}(pp\to X(J/\psi\pi^+\pi^-))=2.5\pm0.7~\mbox{nb},
\label{Prediction-CDF}\ee
which is consistent with the data in Eq.~(\ref{CDF-X3872}). Besides, since the $p_T$ distributions of $X(3872)$ and $\psi'$ production are very similar both for the CMS~\cite{Chatrchyan:2013cld} and D0 data~\cite{D004}, one may expect that the same case would also occur for the CDF data. Therefore, we compare our prediction for the CDF $p_T$ distribution of $X(3872)$, denoted by the red line (central values) and green bands (with errors), with the data of $\psi'$~\cite{CDF09-psi2S} in Fig.~2, where the total cross section of $\psi'$ has been rescaled to be the central values in Eq.~(\ref{Prediction-CDF}). Figure 2 indeed shows a similarity between the measured $p_T$ distribution of $\psi'$ and the predicted one of $X(3872)$, which should be tested by CDF experiment. As for the molecule production,
the predicted total cross section for CDF is too small, which can be obtained by using Eq.~(\ref{Factorization-DD*}) and the matrix element in Eq.~(\ref{fit-molecule}):
\be
\sigma^{molecule}_{CDF}(pp\to X(J/\psi\pi^+\pi^-))=1.1\pm0.4~\mbox{nb},
\ee
which is about 3 times smaller than the CDF data in Eq.~(\ref{CDF-X3872}). This indicates that the CDF data and CMS data are consistent with each other in our model, but they seem to be inharmonious in the molecule model.

\begin{figure}
\includegraphics[width=7cm]{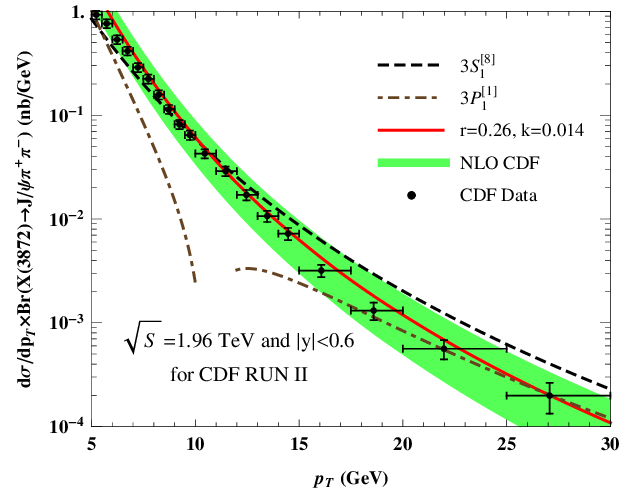}
\caption{\label{fig:diagrams} Comparison between our prediction for the $X(3872)$ $p_T$ distribution, denoted by the red line (central values) and green bands (with errors), and the CDF data for the $\psi'$ $p_T$ distribution~\cite{CDF09-psi2S}. Here the total cross section of the $\psi'$ data has been rescaled to be the central value in Eq.~(\ref{Prediction-CDF}).}
\end{figure}

For the LHCb window ($\sqrt{S}=7~\mbox{TeV},~2.5<y<4.5,~5~\mbox{GeV}<p_T<20~\mbox{GeV}$), using the central values in Eq.~(\ref{r&k}), we predict the total cross section to be
\be
\sigma^{th}_{LHCb}(pp\to X(J/\psi\pi^+\pi^-))=9.4\pm2.2~\mbox{nb},
\label{Prediction-LHCb}\ee
which is about 2 times larger than the experimental data~\cite{LHCb12-CroSec},
\be
\sigma^{ex.}_{tot.}(pp\to X(J/\psi\pi^+\pi^-))=5.4\pm1.5~\mbox{nb}.
\label{LHCb-X3872}\ee
One should note that in Eq.~(\ref{LHCb-X3872}) about $20\%$ of the total cross section comes from B decays; thus, our prediction in Eq.~(\ref{Prediction-LHCb}) is different from the LHCb data by about $2\sigma$ deviation. However, the error bar in Eq.~(\ref{LHCb-X3872}) is larger, and we expect more available data can be used to do the analysis. On the other hand, the total cross section in (\ref{LHCb-X3872}) is dominated by the small-$p_T$ region, i.e., $p_T\sim5~\mbox{GeV}$, with a large $y$ cut. Thus, for one of the initial partons in the protons, the relevant momentum fraction $x$ is very small compared with those of the CMS and CDF windows. Small-$x$ resummation may be needed for improving the theoretical prediction for the LHCb window. But sucha study is beyond the scope of this work, and below we will mainly focus again on the CMS data.

In fact, changing the values of $r$ and $k$ can improve our predictions, especially when the CMS $p_T$ distribution is not sensitive to $r$ (as mentioned above). Thus, we fix $r$ and fit $k$ to the CMS $p_T$ distribution data, and the results are shown in Table~\ref{table1}, where only the central values of $k$ and the predicted cross sections are listed. From the table, one can see that $\sigma^{th}_{CDF}$ and $\sigma^{th}_{LHCb}$ have different $r$ dependences, and thus can not be simultaneously consistent with data in Eqs.~(\ref{CDF-X3872}) and (\ref{LHCb-X3872}).

\begin{table}[htpb]
\caption{ \label{table1} The one-parameter fit to the CMS $p_T$ distribution with fixed $r$. Only the central values of the obtained $k$ and the predicted $\sigma^{th}_{CDF}$ and $\sigma^{th}_{LHCb}$ are listed here.}
\begin{tabular}{|c|c|c|c|c|}\hline
 ~~~~$r$~~~~ & ~~~~$k$~~~~ & $\chi^2/\mbox{d.o.f.}$ & $\sigma^{th}_{CDF}(\mbox{nB})$ & $\sigma^{th}_{LHCb}(\mbox{nB})$ \\\hline
 0.20 & 0.021 & 0.39 & 3.26 & 12.2 \\
 0.25 & 0.015 & 0.17 & 2.63 & 9.87 \\
 0.30 & 0.012 & 0.20 & 2.28 & 8.56 \\
 0.35 & 0.010 & 0.27 & 2.06 & 7.72 \\
 0.40 & 0.008 & 0.34 & 1.90 & 7.14 \\\hline
\end{tabular}
\end{table}

The value of $k$ can also be extracted from the B-decay data, since in our model the branching ratio can be factorized as
\be
Br(B\to X(J/\psi\pi^+\pi^-)K)=Br(B\to\chic K)\cdot k,
\label{Factorization-Bdecay}\ee
and the short-distance branching ratio $Br(B\to\chic K)$ can be extracted by fitting the line shape of the experimental curves~\cite{Zhang2009-fits,Kalashnikova2009-fits}. With a reasonable choice of the decay width of $X(3872)$, the fit in Ref.~\cite{Kalashnikova2009-fits} gives
\be
Br^{fit}(B\to \chic K)=(3.7-5.7)\times 10^{-4},
\label{Br-Btochic2P-fit}\ee
which is consistent with our prediction in Ref.~\cite{Meng-BtoX3872}. By comparing Eqs.~(\ref{Factorization-Bdecay}) and (\ref{Br-Btochic2P-fit}) with
\be
Br(B\to X(J/\psi\pi^+\pi^-)K)=(8.6\pm0.8)\times 10^{-6}~\mbox{\cite{PDG2012}},
\label{Br-BtoX3872-ex}\ee
the constraints on $k$ from B decay are given by
\be
k=0.018\pm0.004,
\label{k-Bdecay}\ee
which is consistent with our fits to the CMS $p_T$ distribution for $r=0.20\mbox{-}0.26$ in Table~\ref{table1}, and thus consistent with the CDF data but not the LHCb data.

As for the value of $|R^{\prime}_{2P}(0)|^2$, if one chooses a larger one than that in Eq.~(\ref{R0}), say, $0.102~\mbox{GeV}^5$~\cite{eichten}, the value of $k$ in Table~\ref{table1} will be decreased by a factor of 0.75 with fixed $r$, and the overlap between Table I and Eq.~(\ref{k-Bdecay}) tends to disappear. Thus, our fits disfavor the larger one.

Finally, with a modest value $Br_0=0.05$ which satisfies the experimental constraints~\cite{PDG2012}, the window of $k$ in Eq.~(\ref{k-Bdecay}) will correspond to the probability of the $\chic$ component in the $X(3872)$:
\be
Z_{c\bar c}=(28\mbox{-}44)\%,
\label{Zcc}\ee
which is consistent both with our original arguments in 2005 (Ref.~\cite{Meng-BtoX3872}) and with the recent fits in Ref.~\cite{Kalashnikova2009-fits}.

\section{Summary}

Within the framework of NRQCD factorization, we evaluate the cross sections of $X(3872)$ at the LHC and Tevatron at NLO in $\alpha_s$ by assuming that the short-distance production proceeds dominantly through its $\chi_{c1}'$ component. The fit of the CMS $p_T$ distribution data~\cite{Chatrchyan:2013cld} gives the ratio $r=0.26\pm0.07$, which is almost the same as that for $\chi_{c1}$~\cite{Chao:chic} and strongly supports the $\chi_{c1}'$-dominated production mechanism for $X(3872)$. The outcomes of the fits can account for the CDF total cross section data~\cite{CDF04,Bignamini2009} and are consistent with the value of $k=Z_{c\bar c}\cdot Br_0$ constrained by the B-decay data simultaneously. The predicted total cross section for the LHCb is larger than the data~\cite{LHCb12-CroSec} by a factor of 2, which may be due to the problem of the fixed-order NRQCD calculation that may not be applicable for the region with small $p_T$ ($p_T\sim 5~\mbox{GeV}$) and large forward rapidity $(2.5<y<4.5)$.   We also evaluate the cross section of the $D^0\bar{D}^{*0}$ molecule~\cite{Braaten2010-X3872HadroPro} at NLO, and find that the molecule-dominated production mechanism for $X(3872)$ seems to be inconsistent with both the $p_T$ distribution and the total cross sections of CMS and CDF.

\begin{acknowledgments}
We thank J.~Z. Li, Y.~Q. Ma, Y.~J. Zhang and G.~Z. Xu for helpful discussions. This work was supported in part by the National Natural
Science Foundation of China (No.~11475005, No.~11075002),
and the National Key Basic Research Program
of China (No.~2015CB856700).
\end{acknowledgments}

\vspace{1.0 cm}

$Note~ added.$ --- While this work was being prepared,another paper~\cite{Butenschoen:2013pxa} appeared to study the prompt production of $X(3872)$ as the $\chi_{c1}'$ meson. Though some of their results are similar to ours, we put stress on the $\chi_{c1}'\mbox{-}D^0\bar{D}^{*0}$ mixing model proposed in Ref.~\cite{Meng-BtoX3872} and find that the $\chi_{c1}'$ component in $X(3872)$ can be dominant if $Br(X\to J/\psi\pi^+\pi^-)<0.04$.

After this work was submitted, there have been a number of experimental and theoretical studies that support our interpretation for the $X(3872)$ being a mixed state of $\chi_{c1}(2P)$ and $DD^*$ components and its production mechanism in hadron collisions.

In 2017, ATLAS measured the $X(3872)$ production cross section in a much larger range of transverse momentum ($10~\mbox{GeV}<p_T<70~\mbox{GeV}$)~\cite{Aaboud:2016vzw}
and found good agreement with our theoretical predictions within the model based on NLO NRQCD, which considers $X(3872)$ to be a mixture of $\chi_{c1}(2P)$ and a $D^0\bar D^{*0}$ molecular state, with the production being dominated by the $\chi_{c1}(2P)$ component~\cite{Aaboud:2016vzw}.

In 2014, LHCb measured\cite{Aaij:2014ala} the ratio $R_{\gamma}=\frac{Br(X(3872)\to \psi'\gamma)}{Br(X(3872)\to J/\psi\gamma)}=(2.46\pm 0.64\pm 0.29$), which more precisely confirmed the earlier BaBar measurement\cite{Aubert:2008ae}. These BaBar and LHCb measurements agree with expectations for a mixture of charmonium and molecular interpretations, but do not support a pure $D^0\bar D^{*0}$ molecular interpretation of the $X(3872)$~\cite{Aaij:2014ala}.

In 2014, the small-$x$ resummation was achieved for $J/\psi$ production at low $p_T$ in $pp(\bar p)$ collisions in Ref.~\cite{Ma:2014mri}. The authors found that the resummation results can match smoothly to those obtained in a NLO collinearly factorized NRQCD formalism such as that used in this work. And the resummation effects can substantially reduce the production cross section of $J/\psi$ predicted by the original NRQCD formalism at $p_T \leq 5~$GeV, especially for the forward region such as the LHCb window. Since the production mechanism of $\chi_{c1}^{(\prime)}$ is similar to that of $J/\psi$, one can expect that deviation between Eqs.~(\ref{Prediction-LHCb}) and (\ref{LHCb-X3872}) might be reduced after the small-$x$ resummation.

The $X(3872)$ was studied by lattice QCD simulations in Ref.~\cite{Prelovsek:2013cra}, where the authors found that the signal of $X(3872)$ can be observed only if both the $c\bar c$ and $DD^*$ operators are included simultaneously in the simulation. This observation, which was unchanged in the later improved lattice simulation where a larger operator basis was used~\cite{Padmanath:2015era}, is a strong support to our proposal that $X(3872)$ is a mixed state of $\chi_{c1}(2P)$ and $DD^*$~\cite{Meng-BtoX3872}.




\end{document}